\documentclass[usenatbib]{emulateapj}
\shortauthors{Cowan \& Agol}
\shorttitle{Albedo and Heat Recirculation on Hot Exoplanets}
\begin{document}

\title{The Statistics of Albedo and Heat Recirculation on Hot Exoplanets}
\author{Nicolas B. Cowan\altaffilmark{1,2},
	 Eric Agol\altaffilmark{2},
	 }
\altaffiltext{1}{CIERA Fellow, Northwestern University, 2131 Tech Dr, Evanston, IL 60208\\
email: n-cowan@northwestern.edu}
\altaffiltext{2}{Astronomy Department, University of Washington,
   Box 351580, Seattle, WA  98195}

\begin{abstract}
If both the day-side and night-side effective temperatures of a planet can be measured, it is possible to estimate its Bond albedo, $0<A_{B}<1$, as well as its day--night heat redistribution efficiency, $0<\varepsilon<1$. We attempt a statistical analysis of the albedo and redistribution efficiency for 24 transiting exoplanets that have at least one published secondary eclipse. For each planet, we show how to calculate a sub-stellar equilibrium temperature, $T_{0}$, and associated uncertainty. We then use a simple model-independent technique to estimate a planet's effective temperature from planet/star flux ratios. We use thermal secondary eclipse measurements ---those obtained at $\lambda > 0.8$~micron--- to estimate day-side effective temperatures, $T_{\rm d}$, and thermal phase variations ---when available--- to estimate night-side effective temperature. We strongly rule out the ``null hypothesis'' of a single $A_{B}$ and $\varepsilon$ for all 24 planets. If we allow each planet to have different parameters, we find that low Bond albedos are favored ($A_{B}<0.35$ at $1\sigma$ confidence), which is an independent confirmation of the low albedos inferred from non-detection of reflected light. Our sample exhibits a wide variety of redistribution efficiencies. When normalized by $T_{0}$, the day-side effective temperatures of the 24 planets describe a uni-modal distribution. The two biggest outliers are GJ~436b (abnormally hot) and HD~80606b (abnormally cool), and these are the only eccentric planets in our sample. The dimensionless quantity $T_{\rm d}/T_{0}$ exhibits no trend with the presence or absence of stratospheric inversions. There is also no clear trend between $T_{\rm d}/T_{0}$ and $T_{0}$. That said, the 6 planets with the greatest sub-stellar equilibrium temperatures ($T>2400$~K) have low $\varepsilon$, as opposed to the 18 cooler planets, which show a variety of recirculation efficiencies.  This hints that the very hottest transiting giant planets are qualitatively different from the merely hot Jupiters. We propose an explanation of this trend based on how a planet's radiative and advective times scale with temperature: both timescales are expected to be shorter for hotter planets, but the temperature-dependance of the radiative timescale is stronger, leading to decreased heat recirculation efficiency. 
\end{abstract}


\section{Introduction}
Short-period exoplanets are expected to have atmospheric compositions and dynamics that differ significantly from Solar System giant planets\footnote{For our purposes a ``short period'' exoplanet is one where the periastron distance is less than $0.1$~AU, regardless of its actual period, and regardless of mass, which may range from Neptune-sized to Brown Dwarf. They are all Class IV and V extrasolar giant planets in the scheme of \cite{Sudarsky_2003}.}. These planets orbit $\sim 100\times$ closer to their host stars than Jupiter does from the Sun.  As a result, they receive $\sim10^4 \times$ more flux and experience tidal forces $\sim10^6\times$  stronger than Jupiter. In contrast to Jupiter, which releases roughly as much power in its interior as it receives from the Sun, short-period exoplanets have power budgets dictated by the flux they receive from their host stars. Roughly speaking, the stellar flux incident on a planet does one of two things: it is reflected back into space, or advected elsewhere on the planet and re-radiated at different wavelengths. The physical parameters that describe these processes are the planet's Bond albedo and redistribution efficiency.

\subsection{Albedo}
Giant planets in the Solar System have albedos greater than 50\% because of the presence of condensed molecules (H$_{2}$O, CH$_{4}$, NH$_{3}$, etc.) in their atmospheres. Planets with effective temperatures exceeding $\sim$400~K should be cloud free, leading to albedos of 0.05--0.4 \citep{Marley_1999}. If pressure-broadened Na and K opacity is important at optical wavelengths \citep[as it is for brown dwarfs,][]{Burrows_2000}, then the Bond albedos of hot Jupiters may be less than 10\% \citep{Sudarsky_2000}. But the very hottest planets, the so-called class V extrasolar giant planets ($T_{\rm eff} > 1500$~K), might have very high albedos due to a high silicate cloud layer \citep{Sudarsky_2000}.  For a planet whose albedo is dominated by clouds (as opposed to Rayleigh scattering) the albedo depends on the composition and size of cloud particles \citep{Seager_2000}.  

Early attempts to observe reflected light from exoplanets \citep{Charbonneau_1999, Cameron_2002, Leigh_2003a, Leigh_2003b, Rodler_2008, Rodler_2010} indicated that they might not be as reflective as Solar System gas giants \citep[for a review, see][]{Langford_2010}. Measurements of HD~209458b taken with the Canadian MOST satellite revealed a very low albedo \citep[$< 8$\%,][]{Rowe_2008}, and it has since been taken for granted that all short-period planets have albedos on par with that of charcoal.
 
From the standpoint of the planet's climate, the important factor is not the albedo at any one wavelength, $A_{\lambda}$, but rather the integrated albedo, weighted by the incident stellar spectrum, known as the Bond albedo and denoted in this paper as $A_{B}$. The relation between $A_{\lambda}$ and the planet's Bond albedo is not trivial. If the albedo is dominated by gray clouds, then the albedo at a single wavelength can indeed be extrapolated to obtain $A_{B}$. For non-gray reflectance spectra, however, it is critical to measure $A_{\lambda}$ at the peak emitting wavelength of the host star to obtain a good estimate of the planet's energy budget.  For example, as pointed out in \cite{Marley_1999}, planets with identical albedo spectra, $A_{\lambda}$, may have radically different $A_{B}$ depending on the spectral type of their host stars.    

\subsection{Redistribution Efficiency}
The first few measurements of hot Jupiter phase variations showed signs that these planets are not all cut from the same cloth.  \cite{Harrington_2006} and \cite{Knutson_2007a} quoted very different phase function amplitudes for the $\upsilon$~Andromeda and HD~189733 systems. It was not clear whether the differences were intrinsic to the planets, however, because the data were taken with different instruments, at different wavelengths, and with very different observation schemes \citep[in any case, subsequent re-analysis of the original data and newly aquired \emph{Spitzer} observations of $\upsilon$~Andromeda~b paint a completely different picture of that system:][]{Crossfield_2010}. 

The uniform study presented in \cite{Cowan_2007}, on the other hand, showed that HD~179949b and HD~209458b exhibit significantly different degrees of heat recirculation, confirming suspicions. But it was not clear whether hot exoplanets were uni-modal or bi-modal in redistribution: are HD~179949b and HD~209458b end-members of a single distribution, or prototypes for two fundamentally different sorts of exoplanets?

The presence or lack of a stratospheric temperature inversion \citep{Hubeny_2003, Fortney_2006b, Burrows_2007b, Burrows_2008, Zahnle_2009} on the day-sides of exoplanets has been invoked to explain a purported bi-modality in recirculation efficiency on hot Jupiters \citep{Fortney_2008}.  The argument, simply put, is that optical absorbers high in the atmosphere of extremely hot Jupiters (equilibrium temperatures greater than $\sim 1700$~K) would absorb incident photons where the radiative timescales are short, making it difficult for these planets to recirculate energy.  The most robust detection of this temperature inversion is for HD~209458b \citep{Knutson_2008}, but this planet does not exhibit a large day-night brightness contrast at 8~$\mu$m \citep{Cowan_2007}. So while temperature inversions seem to exist in the majority of hot Jupiter atmospheres \citep{Knutson_2010}, their connection to circulation efficiency ---if any--- is not clear.

\subsection{Outline of Paper}
It has been suggested \citep[e.g.,][]{Harrington_2006, Cowan_2007} that observations of secondary eclipses and phase variations each constrain a combination of a planet's Bond albedo and circulation efficiency. But observations ---even phase variations--- at a single waveband do little to constrain a planet's energy budget. In this work we show how observations in different wavebands and for different planets can be meaningfully combined to estimate these planetary parameters.

In \S~2 we introduce a simple model to quantify the day-side and night-side energy budget of a short-period planet, and show how a planet's Bond albedo, $A_{B}$, and redistribution efficiency, $\varepsilon$, can be constrained by observations.  In \S~3 we use published observations of 24 transiting planets to estimate day-side and ---where appropriate--- night-side effective temperatures. We construct a two-dimensional distribution function in $A_{B}$ and $\varepsilon$ in \S~4.  We state our conclusions in \S~5. 

\section{Parameterizing the Energy Budget}
\subsection{Incident Flux}
Short-period planets have a power budget entirely dictated by the flux they receive from their host star, which dwarfs tidal heating or remnant heat of formation. Following \cite{Hansen_2008}, we define the equilibrium temperature at the planet's sub-stellar point: $T_{0}(t) = T_{\rm eff} (R_{*}/r(t))^{1/2}$, where $T_{\rm eff}$ and $R_{*}$ are the star's effective temperature and radius, and $r(t)$ is the planet--star distance (for a circular orbit $r$ is simply equal to the semi-major axis, $a$). For shorthand, we define the geometrical factor $a_{*} = a/R_{*}$, which is directly constrained by transit lightcurves \citep{Seager_2003}.

The incident flux on the planet is given by $F_{\rm inc} = \frac{1}{2}\sigma_{B}T_{0}^{4}$, and it is significant that this quantity has some associated uncertainty. For a planet on a circular orbit, the uncertainty in $T_{0}=T_{\rm eff}/\sqrt{a_{*}}$ is related ---to first order--- to the uncertainties in the host star's effective temperature, and the geometrical factor:
\begin{equation}
\frac{\sigma_{T_{0}}^{2}}{T_{0}^{2}} = \frac{\sigma_{T_{\rm eff}}^{2}}{T_{\rm eff}^{2}} + \frac{\sigma_{a_{*}}^{2}}{4a_{*}^{2}}. 
\end{equation}

For a planet with non-zero eccentricity, $T_{0}$ varies with time, but we are only concerned with its value at superior conjunction: secondary eclipse occurs at superior conjunction, when we are seeing the planet's day-side. At that point in the orbit, the planet--star distance is $r_{\rm sc} = a(1-e^{2})/(1-e\sin\omega)$, where $e$ and $\omega$ are the planet's orbital eccentricity and argument of periastron, respectively.
 
For planets with non-zero eccentricity, the uncertainty in $T_{0}$ is given by
\begin{eqnarray}
\frac{\sigma_{T_{0}}^{2}}{T_{0}^{2}} = & \frac{\sigma_{T_{\rm eff}}^{2}}{T_{\rm eff}^{2}} + \frac{\sigma_{a_{*}}^{2}}{4a_{*}^{2}}+\left(\frac{e^{2}\cos^{2}\omega}{1-e^{2}}\right)\sigma_{e\cos\omega}^{2}\cr
& +\left(\frac{e\sin\omega}{1-e^{2}} - \frac{1}{2(1-e\sin\omega)}\right)\sigma_{e\sin\omega}^{2},
\end{eqnarray}
where $\sigma_{e\cos\omega}$ and $\sigma_{e\sin\omega}$ are the observational uncertainties in the two components of the planet's eccentricity\footnote{This formulation is preferable to an error estimate based on $\sigma_{e}$ and $\sigma_{\omega}$, because the eccentricity and argument of periastron are highly correlated in orbital fits.  That said, the uncertainties $\sigma_{e\cos\omega}$ and $\sigma_{e\sin\omega}$ are often not included in the literature, in which case we use a slightly different ---and more conservative--- formulation of the error budget using $\sigma_{e}$ and $\sigma_{\omega}$.}. 

\subsection{Emergent Flux}
At secondary eclipse, and in the absence of albedo or energy circulation, the equilibrium temperature of a region on the planet depends on the normalized projected distance, $\gamma$, from the center of the planetary disc as $T(\gamma) = T_{0} (1-\gamma^{2})^{1/8}$. The thermal secondary eclipse depth in this limit is given by:
\begin{eqnarray}
\label{equilibrium_sed}
\frac{F_{\rm day}}{F_{*}} &= & \left(\frac{R_{p}}{R_{*}} \right)^{2} \left( \frac{h c}{\lambda k T_{0}}\right)^{8}\left(e^{hc/\lambda k T_{\rm b}^{*}}-1\right)\cr
& & \times \int_{0}^{(\lambda k T_{0}/h c)^{8}} \frac{dx}{\exp(x^{-1/8})-1},
\end{eqnarray}
where $T_{b}^{*}$ is the brightness temperature of the star at wavelength $\lambda$. 

In the no-circulation limit, then, the day-side emergent spectrum is not exactly that of a blackbody, even if each annulus has a blackbody spectrum. But these differences are not important for the present study, since we are concerned with bolometric flux. By integrating Equation~\ref{equilibrium_sed} over $\lambda$, one obtains the effective temperature of the day-side in the no-albedo, no-circulation limit: $T_{\varepsilon=0} = (2/3)^{1/4} T_{0}$ \citep[see also][]{Burrows_2008, Hansen_2008}. Indeed, treating the planet's day-side as a uniform hemisphere emitting at this temperature gives nearly the same wavelength dependence as the more complex Equation~\ref{equilibrium_sed}. The $T_{\varepsilon=0}$ temperatures for our sample of 24 transiting planets are shown in Table~1. These set the maximum possible day-side effective temperature we should expect to measure.

The integrated day-side flux in the general ---non-zero circulation--- case is more subtle: heat may be transported to the planet's night-side, and/or to its poles. In this paper we neglect the E-W asymetry in the planet's temperature map due to zonal flows and hence phase offsets in the thermal phase variations. Under this assumption, the day-night temperature contrast can more directly be extracted from the observed thermal phase variations.

In practice, many studies have adopted a single parameter to represent \emph{both} zonal and meridional transport. It is instructive to consider the apparent day-side effective temperatures in various limits: uniform day-side temperature and $T=0$ on the night-side (this is often referred to as the planet's ``equilibrium temperature''): $T_{\rm equ} = (1/2)^{1/4} T_{0}$; in the case of perfect longitudinal transport but no latitudinal transport: $T_{\rm long} = (8/(3\pi^{2}))^{1/4} T_{0}$; and in the limit of a uniform temperature everywhere on the planet: $T_{\rm uni} = (1/4)^{1/4} T_{0}$.

Comparing the apparent day-side temperatures in the three limits of circulation above leads to the following simple parametrization of the day-side effective temperature in terms of the planetary albedo, $A_{B}$, and circulation efficiency, $\varepsilon$:
\begin{equation}
T_{\rm d} = T_{0} (1-A_{B})^{1/4} \left(\frac{2}{3} - \frac{5}{12}\varepsilon \right)^{1/4},
\end{equation}
where $0<\varepsilon<1$. Note that $\varepsilon$ is related to ---but different from--- the $\epsilon$ used in \citep{Cowan_2010}. The former is merely a parametrization of the observed disk-integrated effective temperature, while the latter, which can take values from 0 to $\infty$, is a precisely defined ratio of radiative and advective timescales. The $\epsilon=0$ case is precisely equal to the $\varepsilon=0$ case,  while the $\epsilon\to\infty$ limit is equivalent to $\epsilon\approx 0.95$. 

Our definition of $\varepsilon$ is similar to the \cite{Burrows_2006} definition of $P_{n}$ and yields the same no-circulation limit. But our $\varepsilon=1$ limit produces a lower day-side brightness than the $P_{n}=0.5$ limit, because we assume that the planet's day-side has a uniform temperature distribution in that limit \citep[for a discussion of different redistribution parameterizations, see the appendix of][]{Spiegel_2010}. 

In reality, efficient longitudinal transport (read: fast zonal winds) may lead to more banding and therefore less efficient latitudinal transport. So one could argue that in the limit of perfect day-night temperature homogenization, both the day and night apparent temperatures should be $T_{\rm d}=(8/(3\pi^{2}))^{1/4}T_{0}$, in between the Burrows et al. value of $T_{\rm d}=(1/3)^{1/4}T_{0}$ and that suggested by our parameterization, $T_{\rm d}=(1/4)^{1/4}T_{0}$. At moderate day-night recirculation efficiencies, however, there is a good deal of latitudinal transport (I. Dobbs-Dixon, priv. comm.), so implicitly assuming a constant $T \propto \cos^{1/4}$ latitudinal dependence ---as done by Burrows et al.--- is not founded, either. The bottom line is that any single-parameter implementation of advection is incapable of capturing the real complexities involved, but longitudinal transport is the dominant factor in determining day and night effective temperatures.

Not withstanding the subtleties discussed above and noting that cooling tends to latitudinaly homogenize night-side temperatures \citep{Cowan_2010}, we get a night-side temperature of:
\begin{equation}
T_{\rm n} = T_{0} (1-A_{B})^{1/4} \left(\frac{\varepsilon}{4} \right)^{1/4}.
\end{equation}
Note that $T_{\rm d}$ and $T_{\rm n}$ are the equator-weighted temperatures of their respective hemispheres (ie, as seen by an edge-on viewer).  As such, they will tend to be slightly higher than the hemisphere-averaged temperature, except in the $\varepsilon=1$ limit. This is also why the quantity $T_{\rm d}^{4} + T_{\rm n}^{4}$ is still a weak function of $\varepsilon$. 

\begin{figure}[htb]
\includegraphics[width=84mm]{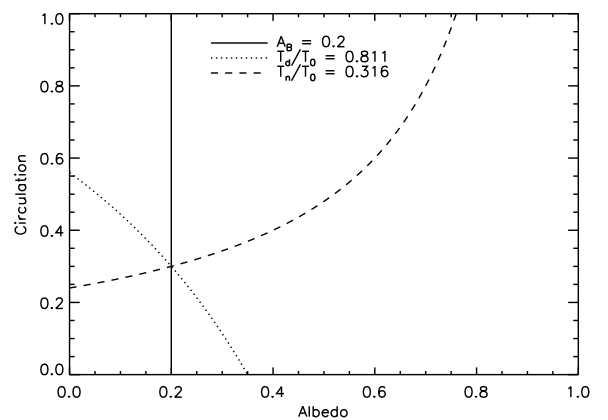}
\caption{Different kinds of idealized observations constrain the Bond albedo, $A_{B}$ and circulation efficiency, $\varepsilon$, differently. A measurement of the secondary eclipse depth at optical wavelengths is a measure of albedo (solid line).  A secondary eclipse depth at thermal wavelengths gives a joint constraint on albedo and recirculation (dotted line). A measurement of the night-side effective temperature from thermal phase variations yields a constraint (the dashed line) nearly orthogonal to the day-side measurement.}
\label{example_exclusion}
\end{figure}

In Figure~\ref{example_exclusion} we show how different kinds of observations constrain $A_{B}$ and $\varepsilon$. For this example, we chose constraints consistent with $A_{B}=0.2$ and $\varepsilon=0.3$. The solid line is a locus of constant $A_{B}$; the dotted line is the locus of constant $T_{d}/T_{0}$; the dashed line is a locus of constant $T_{n}/T_{0}$.  From this figure it is clear that the measurements complement each other: measuring two of the three quantities (Bond albedo, effective day-side or night-side temperatures) uniquely determines the planet's albedo and circulation efficiency. When observations have some associated uncertainty, they define a swath through the $A_{B}$--$\varepsilon$ plane. 

\section{Analysis}
\subsection{Planetary \& Stellar Data}
We begin by considering all the photometric observations of short-period exoplanets published through November 2010, summarized in Table~\ref{observations}. We have discarded photometric observations of non-transiting planets because of their unknown radius and orbital inclination\footnote{For completeness, these are: $\tau$-Bootis~b, $\upsilon$-Andromeda~b, 51~Peg~b, Gl~876d, HD~75289b, HD~179949b and HD~46375b \citep{Charbonneau_1999, Collier_2002, Leigh_2003a, Leigh_2003b, Harrington_2006, Cowan_2007, Seager_2009, Crossfield_2010, Gaulme_2010}}.  This leaves us with 24 transiting exoplanets for which there are observations in at least one waveband at superior conjunction, and in some cases in multiple wavebands and at multiple planetary phases. 

Stellar and planetary data are taken from the Exoplanet Encyclopedia (exoplanet.eu), and references therein. We repeated parts of the analysis with the Exoplanet Data Explorer database (exoplanets.org) and found identical results, within the uncertainties. When the stellar data are not available, we have assumed typical parameters for the appropriate spectral class, and solar metallicity.  Insofar as we are only concerned with the broadband brightnesses of the stars, our results should not depend sensitively on the input stellar parameters. 

Knowing the stars' $T_{\rm eff}$, $\log g$ and [Fe/H], we use the PHOENIX/NextGen stellar spectrum grids \citep{Hauschildt_1999} to determine their brightness temperatures at the observed bandpasses. At each waveband for which eclipse or phase observations have been obtained, we determine the ratio of the stellar flux to the blackbody flux at that grid star's $T_{\rm eff}$.  We then apply this factor to the $T_{\rm eff}$ of the observed star. 

It is worth noting that the choice of stellar model leads to systematic uncertainties in the planetary brightness that are of order the photometric uncertainties.  For example, \cite{Christiansen_2009} use stellar models for HAT-P-7 from \cite{Kurucz_2005}, while we use those of \cite{Hauschildt_1999}. The resulting 8~$\mu$m brightness temperatures for HAT-P-7b differ by as much as 600~K, or slightly more than $1\sigma$. Our uniform use of \cite{Hauschildt_1999} models should alleviate this problem, however.

\subsection{From Flux Ratios to Effective Temperature}
The planet's albedo and recirculation efficiency govern its effective day-side and night-side temperatures, $T_{\rm d}$ and $T_{\rm n}$, respectively.  Observationally, we can only measure the brightness temperature, ideally at a number of different wavelengths: $T_{\rm b}(\lambda)$. If one knew, \emph{a priori}, the emergent spectrum of a planet, one could trivially convert a single brightness temperature to an effective temperature. Alternatively, if observations were obtained at a number of wavelengths bracketing the planet's blackbody peak, it would be possible to estimate the planet's bolometric flux and hence its effective temperature in a model-independent way \citep[e.g.,][]{Barman_2008}.

We adopt the latter empirical approach of converting observed flux ratios into brightness temperatures, then using these to estimate the planet's effective temperature. The secondary eclipse depth in some waveband divided by the transit depth is a direct measure of the ratio of the planet's day-side intensity to the star's intensity at that wavelength, $\psi(\lambda)$. Knowing the star's brightness temperature at a given wavelength, it is possible to compute the apparent brightness temperature of the planet's day side:
\begin{equation}
T_{\rm b}(\lambda) = \frac{h c}{\lambda k} \left[\log\left(1 + \frac{e^{h c/\lambda k T_{b}^{*}(\lambda)} - 1}{\psi(\lambda)}\right)\right]^{-1}.
\end{equation}

On the Rayleigh-Jeans tail, the fractional uncertainty in the brightness temperature is roughly equal to the fractional uncertainty in the eclipse depth; on the Wien tail, the fractional error on brightness temperature can be smaller because the flux is very sensitive to temperature.

By the same token, a secondary eclipse depth and phase variation amplitude at a given wavelength can be combined to obtain a measure of the planet's night-side brightness temperature at that waveband.

Since the albedo and recirculation efficiency of the planet are not known ahead of time, it is not immediately obvious which wavelengths are sensitive to reflected light and which are dominated by thermal emission.  For each planet, we compute the expected blackbody peak if the planet has no albedo and no recirculation of energy, $\lambda_{\varepsilon=0} = 2898/T_{\varepsilon=0}$~$\mu$m. Insofar as real planets will have non-zero albedo and non-zero recirculation, the day side should never reach $T_{\varepsilon=0}$, and the actual spectral energy distribution will peak at slightly longer wavelengths. The coolest planet in our sample, Gl~436b, would exhibit a blackbody peak at $\lambda_{\varepsilon=0}=3.1$~$\mu$m, while the hottest planet we consider, WASP-12b, has $\lambda_{\varepsilon=0}=0.9$~$\mu$m.  In practice this means that ground-based near-IR and space-based mid-IR (e.g., \emph{Spitzer}) observations are assumed to measure thermal emission, while space-based optical observations (MOST, CoRoT, Kepler) may be contaminated by reflected starlight.

\begin{figure}[htb]
\includegraphics[width=84mm]{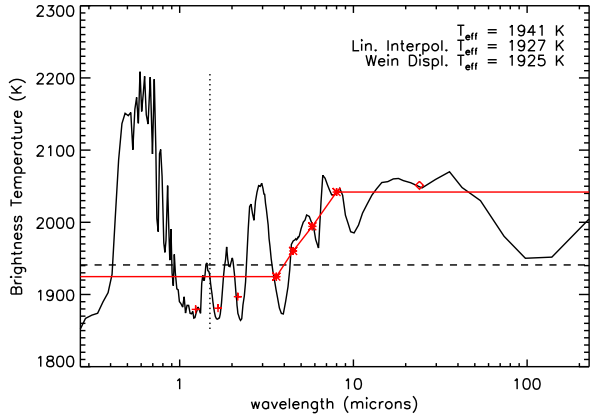}
\caption{The solid black line shows a model spectrum from \cite{Fortney_2008} including only thermal emission (ie: no reflected light).  The planet's effective temperature is shown with the black dashed line, while the expected wavelength of the blackbody peak of the planet is marked with a black dotted line.  The red points show the expected brightness temperatures in the J, H, and K$_{\rm s}$ bands (crosses), as well as the IRAC (asterisks) and MIPS (diamond) instruments on \emph{Spitzer}. The linear interpolation technique described in the text is shown with the red line.}
\label{tres3_2pi_TiO.flx}
\end{figure}

In Figure~\ref{tres3_2pi_TiO.flx}, we demonstrate two alternative techniques to convert an array of brightness temperatures, $T_{\rm b}(\lambda)$, into an estimate of a planet's effective temperature, $T_{\rm eff}$. The solid black line shows a model spectrum of thermal emission from \cite{Fortney_2008}, with an effective temperature of $T_{\rm eff} = 1941$~K shown with the black dashed line. The expected blackbody peak of the planet is marked with a vertical dotted line.  The red points are the expected brightness temperatures in the J, H, and K$_{\rm s}$ bands (crosses), as well as the IRAC (asterisks) and MIPS (diamond) instruments on \emph{Spitzer} \citep{Fazio_2004, Rieke_2004, Werner_2004}. Since the majority of the observations of exoplanets have been obtained with \emph{Spitzer} IRAC, we focus on estimating $T_{\rm eff}$ based \emph{only} on brightness temperatures in those four bandpasses. 

\emph{Wien Displacement:} The first approach is to simply adopt the brightness temperature of the bandpass closest to the planet's blackbody peak (the black dotted line).  If only the four IRAC channels are available, the best one can do is the 3.6~$\mu$m measurement, yielding $T_{\rm eff} = 1925$~K. There is ---however--- some subtlety in estimating the peak wavelength, as this is dependent on knowing the planet's temperature (and hence $A_{B}$ and $\varepsilon$) \emph{a priori}.  

\emph{Linear Interpolation:} The linear interpolation technique, shown with the red line in Figure~\ref{tres3_2pi_TiO.flx}, obviates the need for an estimate of the planet's temperature. The brightness temperature is assumed to be constant shortward of the shortest-$\lambda$ observation, and longward of the longest-$\lambda$ observation.  Between bandpasses, the brightness temperature changes linearly with $\lambda$. As long as the various brightness temperatures do not differ grossly from one another, this technique implicitly gives more weight to observations near the hypothetical blackbody peak. The bolometric flux of this ``model'' spectrum is then computed, and admits a single effective temperature, which is $T_{\rm eff} = 1927$~K for the current example. Since we hope to apply our routine to planets with well sampled blackbody peaks, we adopt the linear interpolation technique, as it can make use of multiple brightness temperature estimates near the peak.

The two techniques described above produce similar effective temperatures, though ---unsurprisingly--- neither gives precisely the correct answer. But these systematic errors are comparable or smaller than the photometric uncertainty in observations of individual brightness temperatures (see Table~1). The best IR observations for the nearest, brightest planetary systems (e.g., HD~189733b and HD~209458b) lead to observational uncertainties of approximately 50~K in brightness temperature.  For many planets, the uncertainty is 100--200~K.  By that metric, either the Wien displacement or the linear interpolation routines give adequate estimates of the effective temperature, with errors of 16~K and 14~K, respectively.  

\begin{figure}[htb]
\includegraphics[width=84mm]{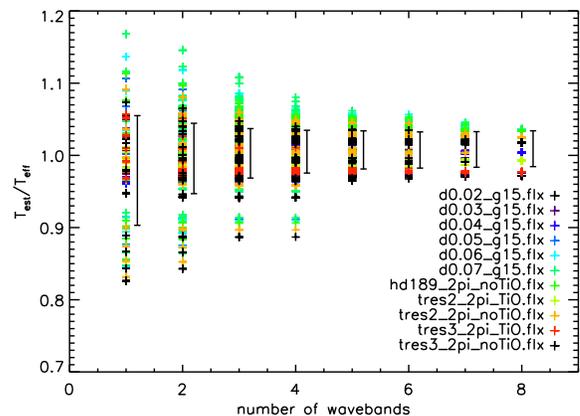}
\caption{The Linear Interpolation technique for estimating day-side effective as tested on a suite of eleven hot Jupiter spectral models provided by J.J.~Fortney. The y-axis shows the estimated day-side effective temperature normalized by the actual model effective temperature.  The x-axis represents the number of brightness temperatures used in the estimate. Each color corresponds to one of the eleven models used in the comparison. The black error bars represent the standard deviation in the normalized temperature estimates.}
\label{T_eff_test}
\end{figure}

We make a more quantitative analysis of the systematic uncertainties involved in the Linear Interpolation temperature estimates as follows. We produce 8800 mock data sets: 100 realizations for 11 models and data in up to 8 wavebands (J, H, K, IRAC, MIPS; Since this numerical experiment chooses random bands from the eight available, the results should not be very different if additional wavebands are considered). We run our Linear Interpolation technique on each of these and plot in Figure~\ref{T_eff_test} the estimated day-side temperature normalized by the actual model effective temperature versus the number of wavebands used in the estimate. The temperature estimates cluster near $T_{\rm est}/T_{\rm eff}=1$, indicating that the technique is not significantly biased. The scatter in estimates decreases as more wavebands are used, from a standard deviation of 7.6\% if only a single brightness temperature is used, down to 2.4\% if photometry is acquired in eight bands. We incorporate this systematic error into our analysis by adding it in quadrature to the observational uncertainties described in the following paragraph. This has the desirable effect that planets with fewer observations have a larger systematic uncertainty on their effective temperature.

In practice, we would like to propagate the photometric uncertainties to the estimate of $T_{\rm eff}$. For the Wien Displacement technique, this uncertainty propagates trivially to the effective temperature. For the linear interpolation technique, a Monte Carlo can be used to estimate the uncertainty in $T_{\rm eff}$: the input eclipse depths are randomly shifted 1000 times in a manner consistent with their photometric uncertainties ---assuming Gaussian errors--- and the effective temperature is recomputed repeatedly. The scatter in the resulting values of $T_{\rm eff}$ provides an estimate of the observational uncertainty in the parameter, to which we add in quadrature the estimate of systematic error described above. The resulting uncertainties are listed in Table~1. These uncertainties should be compared to the uncertainties in $T_{\varepsilon=0}$ (also listed in Table~1), which are computed using the uncertainty in the star's properties and the planet's orbit.   

There are two practical issues with the linear interpolation temperature estimation technique. In some cases, only upper limits have been obtained, therefore one could set $\psi=0$, with the appropriate 1-sigma uncertainty. But this approach leads to huge uncertainties in $T_{\rm eff}$ for planets with a secondary eclipse upper-limit near their blackbody peak. Instead of ``punishing'' these planets, we opt to not use upper-limits (though for completeness we include them in Table~1).  Secondly, when multiple measurements of an eclipse depth have been published for a given waveband, we use the most recent observation, indicated with a superscript ``\emph{e}'' in Table~1. In all cases these observations either explicitly agree with their older counterpart, or agree with the re-analyzed older data.  

\section{Results}
\subsection{Looking for Reflected Light}
For each planet, we use thermal observations (essentially those in the J, H, K$_{s}$, and \emph{Spitzer} bands) to estimate the planet's effective day-side temperature, $T_{\rm d}$, and ---when phase variations are available--- $T_{\rm n}$. These values are listed in Table~1. In five cases (CoRoT-1b, CoRoT-2b, HAT-P-7b, HD~209458b, TrES-2b), secondary eclipses and/or phase variations have been obtained at optical wavelengths. Such observations have the potential to directly constrain the albedo of these planets. One approach is to adopt the $T_{\rm d}$ from thermal observations and calculate the expected contrast ratio at optical wavelengths, under the assumption of blackbody emission \citep[see also][]{Kipping_2010}. Insofar as the observed eclipse depths are deeper than this calculated depth, one can invoke the contribution of reflected light and compute a geometric albedo, $A_{g}$. If one treats the planet as a uniform Lambert sphere, the geometric albedo is related to the spherical albedo at that wavelength by $A_{\lambda} = \frac{3}{2}A_{g}$.  These values are listed in Table~1.

But reflected light is not the only explanation for an unexpectedly deep optical eclipse. Alternatively, the emissivity of the planets may simply be greater at optical wavelengths than at mid-IR wavelengths, in agreement with realistic spectral models of hot Jupiters, which predict brightness temperatures greater than $T_{\rm eff}$ on the Wien tail (see, for example, the Fortney et al. model shown in Figure~\ref{tres3_2pi_TiO.flx}, which does not include reflected light). Note that this increase in emissivity should occur regardless of whether or not the planet has a stratosphere: by definition, the depth at which the optical thermal emission is emitted is the depth at which incident starlight is absorbed, which will necessarily be a hot layer ---assuming the incident stellar spectrum peaks in the optical.

Determining the albedo directly (ie: by observing reflected light) can be difficult for short period planets, because there is no way to distinguish between reflected and re-radiated photons.  The blackbody peaks of the star and planet often differ by less than a micron.  Therefore, unlike Solar System planets, these worlds do not exhibit a minimum in their spectral energy distribution between the reflected and thermal peaks. The hottest ---and therefore most ambiguous case--- of the five transiting planets with optical constraints is HAT-P-7b. If one takes the mid-IR eclipse depths at face value, the planet has a day-side effective temperature of $\sim 2000$~K. When combined with the Kepler observations, one computes an albedo of greater than 50\%. The large day-night amplitude seen in the Kepler bandpass is then simply due to the fact that the planet's night-side reflects no starlight, and the cool day-side can be attributed to high $A_{B}$ and/or $\varepsilon$. If, on the other hand, one takes the optical flux to be entirely thermal in origin ($A_{\lambda}=0$), the day-side effective temperature is $\sim 2800$~K. This is very close to that planet's $T_{\varepsilon=0}$, leaving very little power left for the night-side, again explaining the large day-night contrast observed by Kepler. The truth probably lies somewhere between these two extremes, but in any case this degeneracy will be neatly broken with \emph{Warm Spitzer} observations: the two scenarios outlined above will lead to small and large thermal phase variations, respectively. It is telling that the only optical measurement in Table~1 that is unanimously considered to constrain albedo ---and not thermal emission--- is the MOST observations of HD~209458b \citep{Rowe_2008}, the coolest of the five transiting planets with optical photometric constraints. 

The bottom line is that extracting a constraint on reflected light from optical measurements of hot Jupiters is best done with a detailed spectral model.  But even when reflected light can be directly constrained, converting this constraint on $A_{\lambda}$ into a constraint on $A_{B}$ also requires detailed knowledge of both the star and the planet's spectral energy distributions, making for a model-dependent exercise. 

\subsection{Populating the $A_{B}$-$\varepsilon$ Plane}
Setting aside optical eclipses and direct measurements of albedo, we may use the rich near- and mid-IR data to constrain the Bond albedo and redistribution efficiency of short-period giant planets. We define a $20\times20$ grid in $A_{B}$ and $\varepsilon$ and use Equations~4 \& 5 to calculate the normalized day-side and night-side effective temperatures, $T_{\rm d}/T_{0}$ and $T_{\rm n}/T_{0}$, at each grid point, $(i,j)$. For each planet, we have an observational estimate of the day-side effective temperature, and in three cases we also have an estimate of the night-side effective temperature (as well as associated uncertainties). 

We first verify whether or not the observations are consistent with a single $A_{B}$ and $\varepsilon$. To evaluate this ``null hypothesis'', we compute the usual $\chi^{2} = \sum_{i=1}^{24} ({\rm model}-{\rm data})^{2}/{\rm error}^{2}$ at each grid point.  We use only the estimates of day-side and (when available) night-side effective temperatures to calculate the $\chi^{2}$, giving us 27-2=25 degrees of freedom. The ``best-fit'' has $\chi^{2}=132$ (reduced $\chi^{2}=5.3$), so the current observations strongly rule out a single Bond albedo and redistribution efficiency for all 24 planets. 

For 21 of the 24 planets considered here, we construct a two-dimensional distribution function for each planet as follows:
\begin{equation}
PDF(i,j) = \frac{1}{\sqrt{2\pi\sigma_{\rm d}^{2}}}e^{-(T_{\rm d} - T_{\rm d}(i,j))^{2}/(2\sigma_{\rm d})^{2}}.
\end{equation}
This defines a swath through parameter space with the same shape as the dotted line in Figure~1.

For the three remaining planets (HD~149026b, HD~189733b, HD~209458b), phase variation measurements help break the degeneracy:
\begin{eqnarray}
PDF(i,j) = & \frac{1}{\sqrt{2\pi\sigma_{\rm d}^{2}}}e^{-(T_{\rm d} - T_{\rm d}(i,j))^{2}/(2\sigma_{\rm d})^{2}}\cr 
& \times \frac{1}{\sqrt{2\pi\sigma_{\rm n}^{2}}}e^{-(T_{\rm n} - T_{\rm n}(i,j))^{2}/(2\sigma_{\rm n})^{2}}.
\end{eqnarray}

\begin{figure}[htb]
\includegraphics[width=84mm]{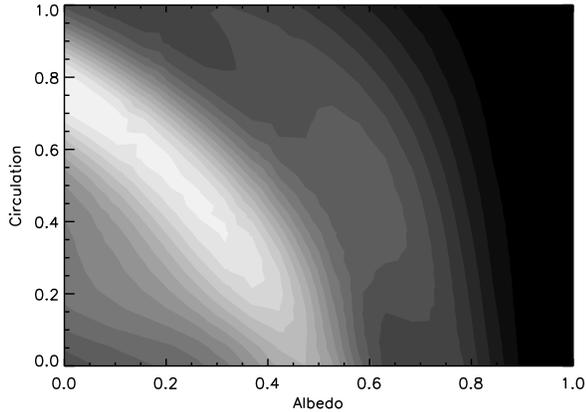}
\caption{The global distribution function for short-period exoplanets in the $A_{B}$--$\varepsilon$ plane.  The gray-scale shows the sum of the normalized probability distribution function for the 24 planets in our sample. The data mostly consist of infrared day-side fluxes, leading to the dominant degeneracy (see first the dotted line in Figure~1).}
\label{pdf_albedo_epsilon}
\end{figure}

We create a two-dimensional normalized probability distribution function (PDF) for each planet, then add these together to create the global PDF shown in Figure~\ref{pdf_albedo_epsilon}. This is a democratic way of representing the data, since each planet's distribution contributes equally.

\begin{figure}[htb]
\includegraphics[width=84mm]{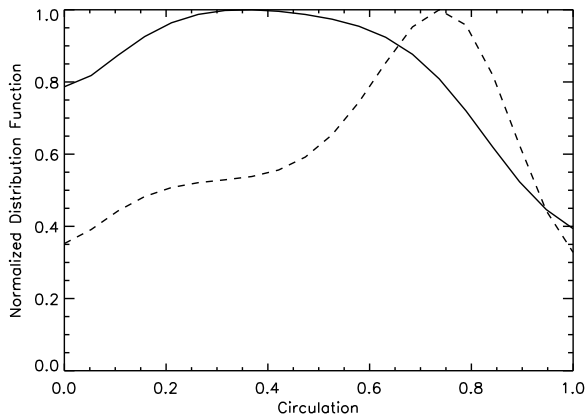}
\caption{The solid black line shows the projection of the 2-dimensional probability distribution function (the gray-scale of Figure~\ref{pdf_albedo_epsilon}) projected onto the $\varepsilon$-axis. The dashed line shows the $\varepsilon$-distribution if one requires that all planets have Bond albedos less than 0.1; under this assumption, we see hints of a bimodal distribution in heat circulation efficiency.}
\label{all_R50_N0_both_circulation_hist}
\end{figure}

\begin{figure}[htb]
\includegraphics[width=84mm]{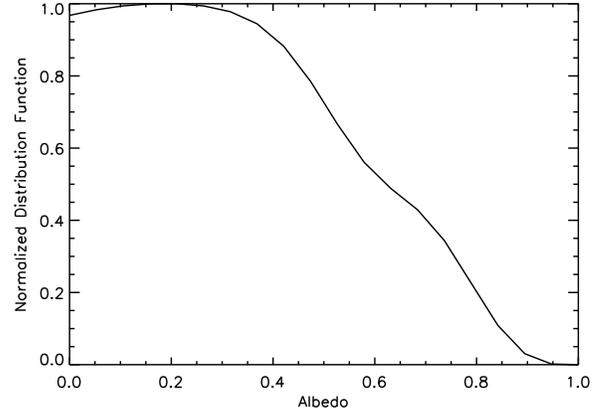}
\caption{The solid black line shows the projection of the 2-dimensional probability distribution function (the gray-scale of Figure~\ref{pdf_albedo_epsilon}) projected onto the $A_{B}$-axis. The cumulative distribution function (not shown) yields a $1\sigma$ upper limit of $A_{B}<0.35$.}
\label{all_R50_N0_both_A_hist}
\end{figure}

In Figures~\ref{all_R50_N0_both_circulation_hist} and \ref{all_R50_N0_both_A_hist} we show the distribution functions for the albedo and circulation of the 24 planets in our sample, obtained by marginalizing the global PDF of Figure~\ref{pdf_albedo_epsilon} over either $A_{B}$ or $\varepsilon$.  

The solid line in Figure~\ref{all_R50_N0_both_circulation_hist} shows no evidence of bi-modality in heat redistribution efficiency, although there is a wide range of behaviors. The dashed line in Figure~\ref{all_R50_N0_both_circulation_hist} shows the $\varepsilon$-distribution if one requires the albedo to be low, $A_{B}<0.1$.  There are then many high-recirculation planets, since advection is the only way to depress the day-side temperature in the absence of albedo. Interestingly, the dashed line \emph{does} show tentative evidence of two separate peaks in $\varepsilon$: if short-period giant planets have uniformly low albedos, then there appear to be two modes of heat recirculation efficiency. We revisit this idea below.

Figure~\ref{all_R50_N0_both_A_hist} shows that planets in this sample are consistent with a low Bond albedo. Note that this constraint is based entirely on near- and mid-infrared observations, and is thus independent from the claims of low albedo based on searches for reflected light \citep[][and references therein]{Rowe_2008}. Furthermore, this is a constraint on the Bond albedo, rather than the albedo in any limited wavelength range. 

\begin{figure}[htb]
\includegraphics[width=84mm]{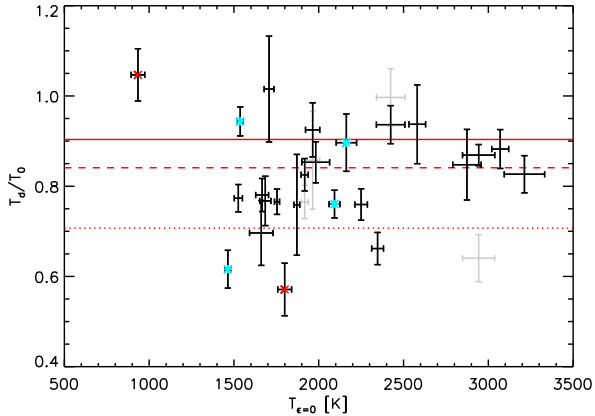}
\caption{The dimensionless day-side effective temperature, $T_{\rm d}/T_{0}$, plotted against the maximum expected day-side temperature, $T_{\varepsilon=0}$. The red lines correspond to three fiducial limits of recirculation, assuming $A_{B}=0$: no recirculation (solid), uniform day-hemisphere (dashed), and uniform planet (dotted). The gray points indicate the default values (using only observations with $\lambda > 0.8$~micron) for the four planets whose optical eclipse depths may be probing thermal emission rather than just reflected light (from left to right: TrES-2b, CoRoT-2b, CoRoT-1b, HAT-P-7b). For these planets we have here elected to include optical measurements in our estimate of the day-side bolometric flux and effective temperature, shown in black.   The cyan asterisks denote those hot Jupiters known \emph{not} to have a stratospheric inversion according to \citep{Knutson_2010}. They are, from left to right: TrES-1b, HD~189733b, TrES-3b, WASP-4b. The two red x's denote the eccentric planets in our sample, which are also the two worst outliers.}
\label{beta_vs_T0}
\end{figure}

In Figure~\ref{beta_vs_T0} we plot the dimensionless day-side effective temperature, $T_{\rm d}/T_{0}$, against the maximum expected day-side temperature, $T_{\varepsilon=0}$. The cyan asterisks in Figure~\ref{beta_vs_T0} show the four hot Jupiters without temperature inversions, while most of the remaining planets have inversions \citep{Knutson_2010}. The presence or absence of an inversion does not appear to affect the efficiency of day--night heat recirculation.

Planets should lie below the solid red line in Figure~\ref{beta_vs_T0}, which denotes $T_{\varepsilon=0}=(2/3)^{1/4}T_{0}$. Of the 24 planets in our sample, only one (Gl~436b) has a day-side effective temperature significantly above the $T_{\varepsilon=0}$ limit\footnote{This is driven by the abnormally high 3.6  micron brightness temperature; including the 4.5 micron eclipse upper limit does not significantly change our estimate of this planet's effective temperature.}. This planet is by far the coolest in our sample, it is on an eccentric orbit, and observations indicate that it may have a non-equilibrium atmosphere \citep{Stevenson_2010}. There is no reason, on the other hand, that planets shouldn't lie below the red dotted line in Figure~\ref{beta_vs_T0}: all it would take is non-zero Bond albedo.  That said, only 3 of the 24 planets we consider are in this region, with the greatest outlier being HD~80606b, a planet on an extremely eccentric orbit with superior conjunction nearly coinciding with periastron. As such, it is likely that much of the energy absorbed by the planet at that point in its orbit performs mechanical work \citep[speeding up winds, puffing up the planet, etc. See also][]{Cowan_2010} rather than merely warming the gas. Gl~436b and HD~80606b are denoted by red x's in Figure~\ref{beta_vs_T0}.

The gray points in Figure~\ref{beta_vs_T0} indicate the default values (using only observations with $\lambda > 0.8$~micron) for the four planets whose optical eclipse depths may be probing thermal emission rather than just reflected light (from left to right: TrES-2b, CoRoT-2b, CoRoT-1b, HAT-P-7b). For these planets we have here elected to use all available flux ratios (including optical observations potentially contaminated by reflected light) to estimate the day-side bolometric flux and effective temperature, shown as black points in Figure~\ref{beta_vs_T0}.

If one takes these day-side effective temperature estimates at face value, it appears that the planets with $T_{\varepsilon=0}<2400$~K exhibit a wide-variety of redistribution efficiencies and/or Bond albedos, but are consistent with $A_{B}=0$. It is worth noting that many of the best characterized planets in this region have $T_{\rm d}/T_{0}\approx 0.75$, and this accounts for the sharp peak in the dotted line of Figure~\ref{all_R50_N0_both_circulation_hist} at $\varepsilon=0.75$. The hottest 6 planets, on the other hand, have uniformly high $T_{\rm d}/T_{0}$, indicating that they have both low Bond albedo \emph{and} low redistribution efficiency. These planets must not have the high-altitude, reflective silicate clouds hypothesized in \cite{Sudarsky_2000}. But this conclusion is dependent on how one interprets the \emph{Kepler} observations of HAT-P-7b: if the large optical flux ratio is due to reflected light, then this planet is cooler than we think, and even the hottest transiting planets exhibit a variety of behaviors.  

\section{Summary \& Conclusions}
We have described how to estimate a planet's incident power budget ($T_{0}$), where the uncertainties are driven by the uncertainties in the host star's effective temperature and size, as well as the planet's orbit. We then described a model-independent technique to estimate the effective temperature of a planet based on planet/star flux ratios obtained at various wavelengths. When the observed day-side and night-side effective temperatures are compared, one can constrain a combination of the planet's Bond albedo, $A_{B}$, and its recirculation efficiency, $\varepsilon$. We applied this analysis on 24 known transiting planets with measured infrared eclipse depths.

Our principal results are:

\noindent 1. Essentially all of the planets are consistent with low Bond albedo.

\noindent 2. We firmly rule out the ``null hypothesis'', whereby all transiting planets can be fit by a single $A_{B}$ and $\varepsilon$. It is not immediately clear whether this stems from differences in Bond albedo, recirculation efficiency, or both. 

\noindent 3. In the few cases where it is possible to unambiguously infer an albedo based on optical eclipse depths, they are extremely low, implying correspondingly low Bond albedos ($<10$\%). If one adopts such low albedos for all the planets in our sample, the discrepancies in day-side effective temperature must be due to differences in recirculation efficiency.  

\noindent 4. These differences in recirculation efficiency do not appear to be correlated with the presence or absence of a stratospheric inversion.

\noindent 5. Planets cooler than $T_{\varepsilon=0}=2400$~K exhibit a wide variety of circulation efficiencies that do not appear to be correlated with equilibrium temperature. Alternatively, these planets may have different (but generally low) albedos. Planets hotter than $T_{\varepsilon=0}=2400$~K have uniformly low redistribution efficiencies and albedos.

The apparent decrease in advective efficiency with increasing planetary temperature remains unexplained. One hypothesis, mentioned earlier, is that TiO and VO would provide additional optical opacity in atmospheres hotter than $T\sim1700$~K, leading to temperature inversions and reduced heat recirculation on these planets \citep{Fortney_2008}. But if our sample shows any sharp change it behavior it occurs near 2400~K, rather than ~1700~K.  One could invoke another optical absorber, but in any case the lack of correlation ---pointed out in this work and elsewhere--- between the presence of a temperature inversion and the efficiency of heat recirculation makes this explanation suspect. Another possible explanation for the observed trend is that the hottest planets have the most ionized atmospheres and may suffer the most severe magnetic drag \citep{Perna_2010}.

The simplest explanation for this trend is simply that the radiative time is a steeper function of temperature than the advective time: advective efficiency is given roughly by the ratio of the radiative and advective times \citep[eg:][]{Cowan_2010}. In the limit of Newtonian cooling, the radiative time scales as $\tau_{\rm rad} \propto T^{-3}$. If one assumes the wind speed to be of order the local sound speed, then the advective time scales as $\tau_{\rm adv} \propto T^{-0.5}$. One might therefore naively expect the advective efficiency to scale as $T^{-2.5}$. Such an explanation would not explain the apparent sharp transition seen at 2400~K, however.   

The combination of optical observations of secondary eclipses and thermal observations of phase variations is the best way to constrain planetary albedo and circulation. The optical observations should be taken near the star's blackbody peak, both to maximize signal-to-noise, and to avoid contamination from the planet's thermal emission, but this separation may not be possible for the hottest transiting planets. The thermal observations, likewise, should be near the planet's blackbody peak to better constrain its bolometric flux.  Note that this wavelength is shortward of the ideal contrast ratio, which typically falls on the planet's Rayleigh-Jeans tail. Furthermore, the thermal phase observations should span a full planetary orbit: the light curve minimum is the most sensitive measure of $\varepsilon$, and should occur nearly half an orbit apart from the light curve maximum, despite skewed diurnal heating patterns \citep{Cowan_2008, Cowan_2010}.  This means that observing campaigns that only cover a little more than half an orbit (transit $\to$ eclipse) are probably underestimating the real peak-trough phase amplitude. 

A possible improvement to this study would be to perform a uniform data reduction for all the \emph{Spitzer} exoplanet observations of hot Jupiters.  These data make up the majority of the constraints presented in our study and most are publicly available. And while the published observations were analyzed in disparate ways, a consensus approach to correcting detector systematics is beginning to emerge.

\acknowledgments
N.B.C. acknowledges useful discussions of aspects of this work with T. Robinson, M.S. Marley, J.J. Fortney, T.S. Barman and D.S. Spiegel.  Thanks to our referee B.M.S. Hansen for insightful feedback, and to E.D. Feigelson for suggestions about statistical methods. N.B.C. was supported by the Natural Sciences and Engineering Research Council of Canada. E.A. is supported by a National Science Foundation Career Grant. Support for this work was provided by NASA through an award issued by JPL/Caltech. This research has made use of the Exoplanet Orbit Database and the Exoplanet Data Explorer at exoplanets.org.

\clearpage
\LongTables

\begin{deluxetable*}{lclrrr|r}
\tabletypesize{\scriptsize}
\tablecaption{Secondary Eclipses \& Phase Variations of Exoplanets \label{observations}}
\tablewidth{0pt}
\tablehead{
\colhead{Planet} & \colhead{$T_{\varepsilon=0}$ [K]$^{a}$} & \colhead{$\lambda$ [$\mu$m]$^{b}$} & \colhead{Eclipse Depth$^{c}$} & $T_{\rm bright}$ [K]&\colhead{Phase Amplitude$^{c}$} & \colhead{Derived Quantities$^{d}$}}
\startdata
CoRoT-1b$^{1}$ & 2424(84) & 0.60(0.42) & $1.6(6)\times10^{-4}$ & 2726(141)& &$T_{\rm d}$=2674(144) K   \\
&  & 0.71(0.25) & $1.26(33)\times10^{-4}$  & 2409(75)& $1.0(3)\times10^{-4}$ &$A_{\lambda} <0.1$    \\
&  & 2.10(0.02) & $2.8(5)\times10^{-3}$  & 2741(125)& & $T_{\rm d}(A=0)$=2515(84) K   \\
&  & 2.15(0.32) & $3.36(42)\times10^{-3}$  & 2490(157)& &   \\
&  & 3.6(0.75) & $4.15(42)\times10^{-3}$ & 2098(116)& & \\
&  & 4.5(1.0) & $4.82(42)\times10^{-3}$&2084(106)&\\
\hline
CoRoT-2b$^{2}$ & 1964(42) & 0.60(0.42) & $6(2)\times10^{-5}$  & 2315(85)& &$T_{\rm d}$=1864(233) K \\
&  &  0.71(0.25) & $1.02(20)\times10^{-4}$  & 2215(49)& & $A_{\lambda} = 0.16(7)$ \\
&  &  1.65(0.25) & $<1.7\times10^{-3}$ (3$\sigma$)  & & &$T_{\rm d}(A=0)$=2010(144) K    \\
&  &  2.15(0.32) & $1.6(9)\times10^{-3}$  & 1914(292)& &    \\
&  & 3.6(0.75) & $3.55(20)\times10^{-3}$ & 1798(40)& & \\
&  & 4.5(1.0)$^{e}$ & $4.75(19)\times10^{-3}$& 1791(33)&&\\
&  & 4.5(1.0) & $5.10(42)\times10^{-3}$ & & &      \\
&  & 8.0(2.9) & $4.1(1.1)\times10^{-3}$ & & &      \\
&  & 8.0(2.9)$^{e}$ & $4.09(80)\times10^{-3}$ & 1318(143)&  &\\
\hline
Gl~436b$^{3}$ & 934(41)&3.6(0.75)& $4.1(3)\times10^{-4}$ & 1145(23)& &$T_{\rm d}$=1082(38) K \\
&   &  4.5(1.0)&$<1.0\times10^{-4}$ (3$\sigma$)& &&\\
&   &  5.8(1.4)&$3.3(1.4)\times10^{-4}$&797(106) &&\\
&   &  8.0(2.9)$^{e}$&$4.52(27)\times10^{-4}$& 737(17)&&\\
&   &  8.0(2.9) & $5.7(8)\times10^{-4}$ & &     & \\
&   &  8.0(2.9) & $5.4(7)\times10^{-4}$  & &    & \\
&   &  16(5) & $1.40(27)\times10^{-3}$  &963(126) &    & \\
&   &  24(9) & $1.75(41)\times10^{-3}$  &1016(182) &    & \\
\hline
HAT-P-1b$^{4}$ & 1666(38)& 3.6(0.75) & $8.0(8)\times10^{-4}$ &1420(47) &  &$T_{\rm d}$=1439(59) K   \\
 & &   4.5(1.0) & $1.35(22)\times10^{-3}$  & 1507(100)& &    \\
 & &   5.8(1.4) & $2.03(31)\times10^{-3}$  &1626(128) & &    \\
 & &   8.0(2.9) & $2.38(40)\times10^{-3}$  &1564(151) & &    \\
\hline
HAT-P-7b$^{5}$ & 2943(95)& 0.65(0.4) & $1.30(11)\times10^{-4}$  &3037(35) & $1.22(16)\times10^{-4}$&$T_{\rm d}$=2086(156) K \\
&    &  3.6(0.75) & $9.8(1.7)\times10^{-4}$  & 2063(152)& & $A_{\lambda} = 0.58(5)$ \\
&   &  4.5(1.0) & $1.59(22)\times10^{-3}$  &2378(179) & &  $T_{\rm d}(A=0)$=2830(86) K    \\
&    & 5.8(1.4) & $2.45(31)\times10^{-3}$  &2851(235) & &    \\
&   &  8.0(2.9) & $2.25(52)\times10^{-3}$  &2512(403) & &     \\
\hline
HD~80606b$^{6}$ & 1799(50)& 8.0(2.9) & $1.36(18)\times10^{-3}$  &1137(73) & & $T_{\rm d}$=1137(113) K \\
\hline
HD~149026b$^{7}$ & 1871(17) & 8.0(2.9)$^{e}$ & $3.7(0.8)\times10^{-4}$ &976(276) & $2.3(7)\times10^{-4}$& $T_{\rm d}$=1571(231) K \\
&  & 8.0(2.9) & $8.4(1.1)\times10^{-4}$ & &  &$T_{\rm n}$=976(286) K  \\
\hline
HD~189733b$^{8}$ & 1537(16)& 2.15(32)  &$<4.0\times10^{-4} (1\sigma)$& &&$T_{\rm d}$=1605(52) K  \\
&   & 3.6(0.75) & $2.56(14)\times10^{-3}$&1639(34) &  &  $T_{\rm n}$=1107(132) K\\
&   &  4.5(1.0) & $2.14(20)\times10^{-3}$ &1318(45) & &  \\
&   &  5.8(1.4) & $3.10(34)\times10^{-3}$  &1368(69) & &    \\
&    &  8.0(2.9) & $3.381(55)\times10^{-3}$  & & &    \\
&    &  8.0(2.9) & $3.91(22)\times10^{-3}$  & &$1.2(2)\times10^{-3}$ &   \\
&    &  8.0(2.9)$^{e}$ & $3.440(36)\times10^{-3}$  &1259(7) &$1.2(4)\times10^{-3}$ &    \\
&    &  16(5) & $5.51(30)\times10^{-3}$  & 1338(52)& &    \\
&    &  24(9) & $5.98(38)\times10^{-3}$  & & &    \\
&    &  24(9)$^{e}$ & $5.36(27)\times10^{-3}$ &1202(46) & $1.3(3)\times10^{-3}$ &    \\
\hline
HD~209458b$^{9}$ & 1754(15)& 0.5(0.3) & $7(9)\times10^{-6}$&2368(156)  & &$T_{\rm d}$=1486(53) K   \\
&   &  2.15(0.32) & $<3\times10^{-4} (1\sigma)$   & & &$A_{\lambda} = 0.09(7)$  \\
&    &  3.6(0.75) & $9.4(9)\times10^{-4}$  &1446(45) & & $T_{\rm d}(A=0)$=2031(128) K   \\
&   &  4.5(1.0) & $2.13(15)\times10^{-3}$  &1757(57) & & $T_{\rm n}$=1476(304) K   \\
&    & 5.8(1.4) & $3.01(43)\times10^{-3}$  &1890(149) & &    \\
&   &  8.0(2.9) & $2.40(26)\times10^{-3}$  &1480(94) &$<1.5\times10^{-3}$ ($2\sigma$)  & \\
&   &  24(9) & $2.60(44)\times10^{-3}$  & 1131(143)& &    \\
\hline
OGLE-TR-56b$^{10}$ & 2874(84)& 0.90(0.15) & $3.63(91)\times10^{-4}$ &2696(116)  & & $T_{\rm d}$=2696(236) K \\
\hline
OGLE-TR-113b$^{11}$ & 1716(33)& 2.15(0.32) & $1.7(5)\times10^{-3}$& 1918(164)  & & $T_{\rm d}$=1918(219) K \\
\hline
TrES-1b$^{12}$ & 1464(16) & 3.6(0.75) & $< 1.5\times10^{-3} (1\sigma)$&   & & $T_{\rm d}$=998(67) K  \\
& &  4.5(1.0) & $6.6(1.3)\times10^{-4}$ &972(56)  & &    \\
&     & 8.0(2.9) & $2.25(36)\times10^{-3}$& 1152(94)  &   &  \\
\hline
TrES-2b$^{13}$ & 1917(21)& 0.65(0.4) & $1.14(78)\times10^{-5}$ &2020(132)&  &$T_{\rm d}$=1623(76) K\\
&     & 2.15(0.32) & $6.2(1.2)\times10^{-4}$ &1655(80)&  &$A_{\lambda} = 0.06(3)$\\
&     & 3.6(0.75) & $1.27(21)\times10^{-3}$  &1490(84)&  &$T_{\rm d}(A=0) = 1751(80)$~K   \\
&     & 4.5(1.0) & $2.30(24)\times10^{-3}$  &1652(74) &&    \\
&     & 5.8(1.4) & $1.99(54)\times10^{-3}$  &1373(177) & &   \\
&     & 8.0(2.9) & $3.59(60)\times10^{-3}$  &1659(163) & &   \\
\hline
TrES-3b$^{14}$ & 2093(32)& 0.7(0.3) & $<6.2\times10^{-4}$ ($1\sigma$) &&  &$T_{\rm d}$=1761(66) K \\
&     & 1.25(0.16) & $<5.1\times10^{-4}$ ($3\sigma$) &&  & \\
&     & 2.15(0.32) & $2.41(43)\times10^{-3}$ &  & & \\
&     & 2.15(0.32)$^{e}$ & $1.33(17)\times10^{-3}$ &1770(58) & & \\
&     & 3.6(0.75) & $3.46(35)\times10^{-3}$  &1818(73)   & & \\
&     & 4.5(1.0) & $3.72(54)\times10^{-3}$  &1649(107)   & & \\
&     & 5.8(1.4) & $4.49(97)\times10^{-3}$  &1621(173)   & & \\
&     & 8.0(2.9)  &  $4.75(46)\times10^{-3}$   &1480(82)  & & \\
\hline
TrES-4b$^{15}$  & 2250(37)& 3.6(0.75) & $1.37(11)\times10^{-3}$ &1889(63) & &$T_{\rm d}$=1891(81) K  \\
&      & 4.5(1.0) & $1.48(16)\times10^{-3}$   &1727(83)  &&  \\
&      & 5.8(1.4) & $2.61(59)\times10^{-3}$   &2112(283)  & & \\
&     & 8.0(2.9) & $3.18(44)\times10^{-3}$   & 2168(197) &  &\\
\hline
WASP-1b$^{16}$  & 2347(35)& 3.6(0.75) & $1.17(16)\times10^{-3}$ &1678(87)&  &$T_{\rm d}$=1719(89) K  \\
&      & 4.5(1.0) & $2.12(21)\times10^{-3}$   &1923(91)  &  &\\
&      & 5.8(1.4) & $2.82(60)\times10^{-3}$  &2042(253)  &  &\\
&     & 8.0(2.9) & $4.70(46)\times10^{-3}$   &2587(176)  &  &\\
\hline
WASP-2b$^{17}$  & 1661(69)& 3.6(0.75) & $8.3(3.5)\times10^{-4}$ &1264(164)&  &$T_{\rm d}$=1280(121) K  \\
&      & 4.5(1.0) & $1.69(17)\times10^{-3}$   &1380(53)  &  &\\
&      & 5.8(1.4) & $1.92(77)\times10^{-3}$   &1299(232)  &  &\\
&     & 8.0(2.9) & $2.85(59)\times10^{-3}$   &1372(154)  &  &\\
\hline
WASP-4b$^{18}$  & 2163(60)& 3.6(0.75) & $3.19(31)\times10^{-3}$ &2156(97)&  &$T_{\rm d}$=2146(140) K  \\
&      & 4.5(1.0) & $3.43(27)\times10^{-3}$    &1971(75) & &\\
\hline
WASP-12b$^{19}$ & 3213(119) & 0.9(0.15) & $8.2(1.5)\times10^{-4}$ &3002(104) & &$T_{\rm d}$=2939(98) K \\ 
& & 1.25(0.16)& $1.31(28)\times10^{-3}$&2894(149) &&\\
&   & 1.65(0.25)& $1.76(18)\times10^{-3}$ &2823(88)&&\\
&   & 2.15(0.32)& $3.09(13)\times10^{-3}$ &3018(51)& & \\
&   & 3.6(0.75) & $3.79(13)\times10^{-3}$&2704(49) &&\\
&   & 4.5(1.0) & $3.82(19)\times10^{-3}$  &2486(68)&&\\
&   & 5.8(1.4) & $6.29(52)\times10^{-3}$&3167(179) &&\\
&   & 8.0(2.9) & $6.36(67)\times10^{-3}$  &2996(229)&&\\
\hline
WASP-18b$^{20}$ & 3070(50) & 3.6(0.75) & $3.1(2)\times10^{-3}$ &3000(107)& &$T_{\rm d}$=2998(138) K\\
&   & 4.5(1.0) & $3.8(3)\times10^{-3}$  &3128(150) &&\\
&   & 5.8(1.4) & $4.1(2)\times10^{-3}$ &3095(103) &&\\
&   & 8.0(2.9) & $4.3(3)\times10^{-3}$ &2991(153) && \\
\hline
WASP-19b$^{21}$ & 2581(49)& 1.65(0.25) & $2.59(45)\times10^{-3}$&2677(135)& &$T_{\rm d}$=2677(244) K\\
\hline
XO-1b$^{22}$ & 1526(24) & 3.6(0.75) & $8.6(7)\times10^{-4}$ &1300(32)  &&$T_{\rm d}$=1306(47) K \\
&  &   4.5(1.0) & $1.22(9)\times10^{-3}$   &1265(34)  & & \\
&   &   5.8(1.4) & $2.61(31)\times10^{-3}$   &1546(89)  &&  \\
&   &   8.0(2.9) & $2.10(29)\times10^{-3}$  &1211(87)   & & \\
\hline
XO-2$^{23}$ & 1685(33) & 3.6(0.75) & $8.1(1.7)\times10^{-4}$  &1447(102)& &$T_{\rm d}$=1431(98) K \\
& &   4.5(1.0) & $9.8(2.0)\times10^{-4}$  & 1341(105)& &   \\
& &   5.8(1.4) & $1.67(36)\times10^{-3}$  &1497(155) &  &  \\
& &   8.0(2.9) & $1.33(49)\times10^{-3}$  &1179(219) &   & \\
\hline
XO-3$^{24}$ & 1982(82) & 3.6(0.75) & $1.01(4)\times10^{-3}$  &1875(30) && $T_{\rm d}$=1871(63) K \\
& &   4.5(1.0) & $1.43(6)\times10^{-3}$  &1965(40) & &   \\
& &   5.8(1.4) & $1.34(49)\times10^{-3}$  &1716(330) & &   \\
& &   8.0(2.9) & $1.50(36)\times10^{-3}$  &1625(236) &  &  \\
\enddata
\tablenotetext{a}{The planet's expected day-side effective temperature in the absence of reflection or recirculation ($A_{B}=0$, $\varepsilon=0$). The $1\sigma$ uncertainty is shown in parenthese.}
\tablenotetext{b}{The bandwidth is shown in parenthese.}
\tablenotetext{c}{Eclipse depths and phase amplitudes are unitless, since they are measured relative to stellar flux.}
\tablenotetext{d}{$T_{\rm d}$ and $T_{\rm n}$ denote the day-side and night-side effective temperatures of the planet, as estimated from thermal secondary eclipse depths and thermal phase variations, respectively. The estimated $1\sigma$ uncertainties are shown in parentheses. The default day-side temperature is computed using only observations at $\lambda > 0.8$~$\mu$m. Eclipse measurements at shorter wavelengths may then be used to estimate the planet's albedo at those wavelengths, $A_{\lambda}$. Note that this is a spherical albedo; the geometric albedo is given by $A_{g} = \frac{2}{3}A_{\lambda}$. If ---on the other hand--- $A_{B}=0$ is assumed, then all the day-side flux is thermal, regardless of waveband, yielding the second $T_{\rm d}$ estimate.}
\tablenotetext{e}{When multiple measurements of an eclipse depth have been published in a given waveband, we use the most recent observation. In all cases these observations are either explicitly agree with their older counterpart, or agree with the re-analyzed older data.}
\tablenotetext{1}{\cite{Snellen_2009a, Alonso_2009a, Gillon_2009a, Rogers_2009, Deming_2010}, $^{2}$\cite{Alonso_2009b, Snellen_2010, Gillon_2009b, Alonso_2010, Deming_2010}, $^{3}$\cite{Deming_2007, Demory_2007, Stevenson_2010}; Knutson et al.\ in prep., $^{4}$\cite{Todorov_2009}, $^{5}$\cite{Borucki_2009, Christiansen_2009}, $^{6}$\cite{Laughlin_2009}, $^{7}$\cite{Knutson_2009c}, $^{8}$\cite{Deming_2006, Knutson_2007a, Barnes_2007, Charbonneau_2008, Knutson_2009a, Agol_2010}, $^{9}$\cite{Richardson_2003, Deming_2005, Cowan_2007, Rowe_2008, Knutson_2008}, $^{10}$\cite{Sing_2009}, $^{11}$\cite{Snellen_2007}, $^{12}$\cite{Charbonneau_2005, Knutson_2007b}, $^{13}$\cite{ODonovan_2010, Croll_2010a, Kipping_2010b}, $^{14}$\cite{Fressin_2009, Croll_2010b, Christiansen_2010b}, $^{15}$\cite{Knutson_2009b}, $^{16,17}$\cite{Wheatley_2010}, $^{18}$\cite{Beerer_2010}, $^{19}$\cite{Lopez-Morales_2010, Campo_2010, Croll_2010c}, $^{20}$\cite{Nymeyer_2010}, $^{21}$\cite{Anderson_2010a}, $^{22}$\cite{Machalek_2008}, $^{23}$\cite{Machalek_2009}, $^{24}$\cite{Machalek_2010}}
\end{deluxetable*}

\end{document}